# The partial asteroidal occultation of Betelgeuse on Jan 2, 2012.

**Costantino Sigismondi**

Galileo Ferraris Institute and International Center for Relativistic Astrophysics, Rome, Italy

sigismondi@icra.it

**Abstract:** The asteroid (147857) 2005 UW381 will pass over the supergiant star Betelgeuse on January 2$^{nd}$ 2012. The event is visible on a limited geographical region, and the magnitude drop is only 0.01 magnitudes for a maximum duration of 3.6 seconds. The opportunity to measure this phenomenon can be interesting for dealing with extrasolar planetary transits.

**The occultation:**

The data on the partial asteroidal occultation are presented in the image available at the web link

http://www.asteroidoccultation.com/2012_01/0102_147857_29098_Map.gif

We can to do the following considerations: assuming that Betelgeuse has a diameter of 31 mas (as indicated in that website) and the asteroid has a diameter of 3 mas.

The ratio between the areas is $(3/31)^2$ and the luminosity of Betelgeuse will be

$[1-(3/31)^2]*100\% = 99.06\%$ of its maximum

In terms of magnitude the difference according to the Pogson's law the decrease of the magnitude of the star will be

$2.5 \cdot \log(0.9906) = -0.0102$ magnitudes

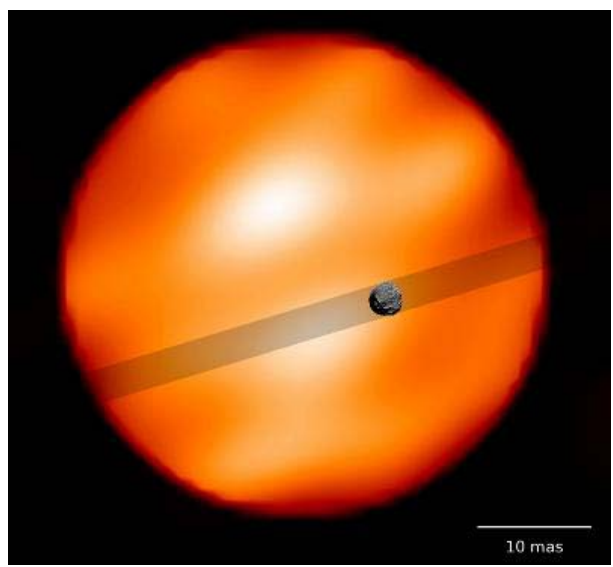

Fig. 1 A sketch of the phenomenon made by Alfons Gabel (IOTA/ES).

The extension of the stellar atmosphere stellar could be measured by timing that passage from different locations along the predicted path of visibility of that transit.

Since the occulting asteroid has a parallax of 4.648" it means that this is the Earth's radius seen from its position. The following equivalence 1.37 km = 1 mas implies that the 31 mas diameter of Betelgeuse corresponds to 42.5 km on Earth, where the asteroid appears over the disk, plus 4.1 km North and 4.1 km South of the central path, where the asteroid is grazing the star's disk.

The locations of the central path, with Northern and Southern limits are plotted on the website of IOTA/ES http://www.asteroidoccultation.com/2012_01/0102_147857_29098_Summary.txt

**The Star**

Betelgeuse is the alpha of Orion. The angular diameter is the largest observed for a star, and it is the first star of which some surface's details have been observed.

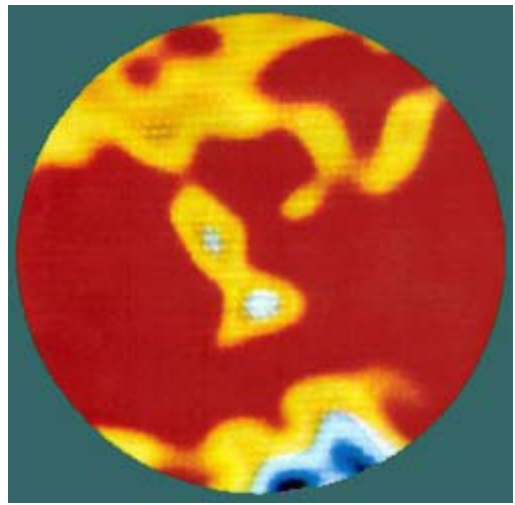

Fig. 2 With the speckle interferometry technique the surface of Betelgeuse has been inspected at the Kitt Peak Observatory.

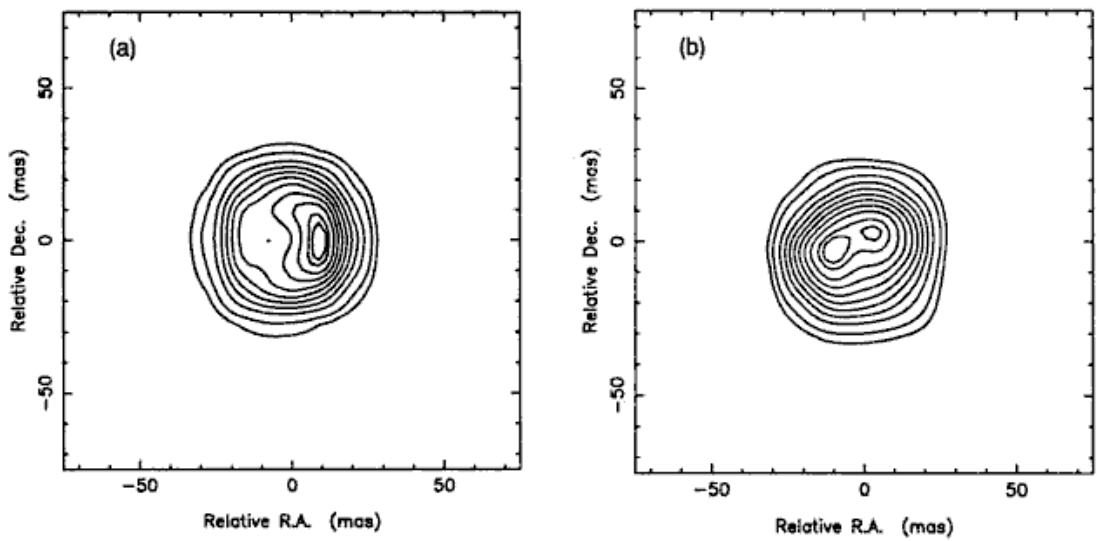

Fig. 3 Betelgeuse at the William Herschel 4.2 m Telescope – Ground-based High Resolution Imaging Laboratory (La Palma), from [1]. Left (a): reconstruction of Betelgeuse with Maximum Entropy Method at 710 nm in February 1989 and Right (b): in January 1991.
Contours are 5, 10, 20… 90, 95 of the peak intensity.

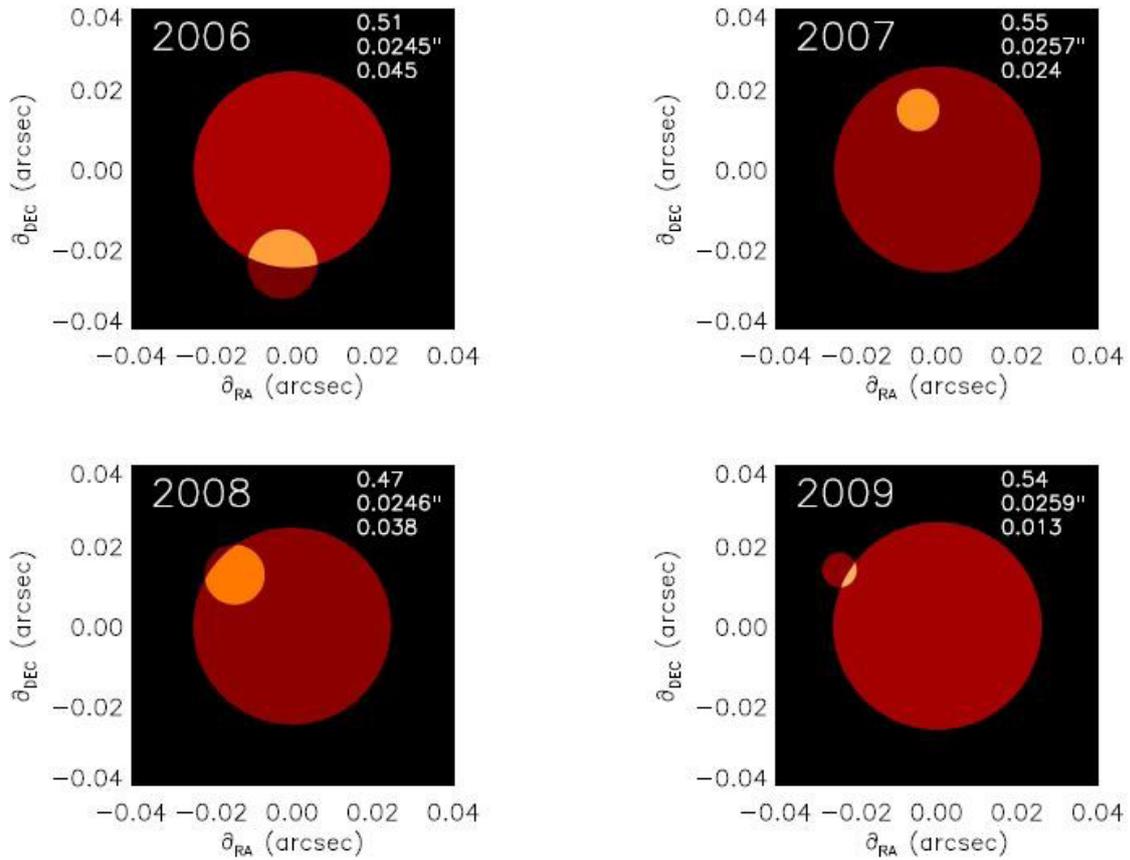

Fig. 4 Image from the paper "The many faces of Betelgeuse".[2] A geometrical model for an hot spot, calculated from different measures made on Betelgeuse at the Berkeley Infrared Space Interferometer of 11.15μm in 2006, 2007, 2008 and 2009. Each figure includes the fit parameters; the fraction of the total flux from the star, the stella radius in arcseconds, and the fraction of the total flux from the point. The point sources have been give the uniform disk sizes that they would have if they represented regions at a temperature of 7200 K. The upper limit on the point source diameter is 20 mas.

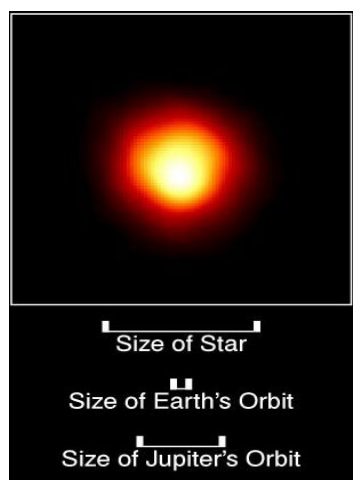

Fig. 5 The angular diameter of Betelgeuse with the Hubble Space Telescope.[3]

In the UV the stellar disk at 255 nm is 2.5 times larger than in visible light. The atmosphere of that supergiant star is as large as 125 mas. Already Michelson and Pease [4] considered the limb darkening reducing the measured diameter.

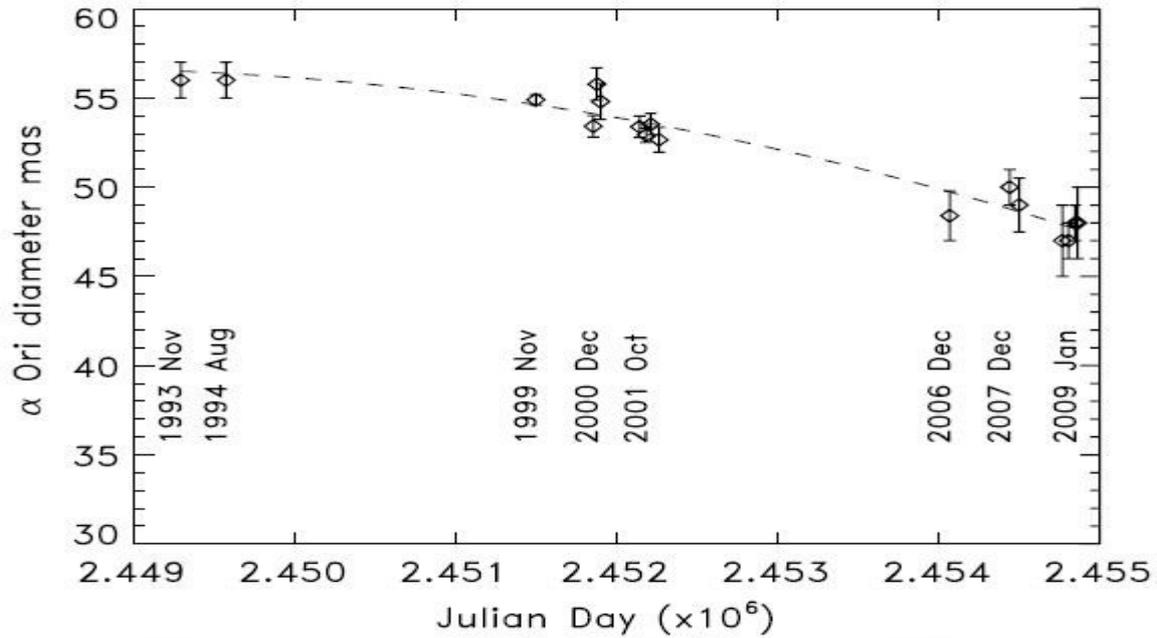

Fig. 6 The diameter of α Orionis measured at 11.15μm by the Betelgeuse at the Berkeley ISI Infrared Space Interferometer over 15 years. Values previously published are quoted in Townes et al., [5] were this figure has been taken. The dashed line is a quadratic fit to the data.

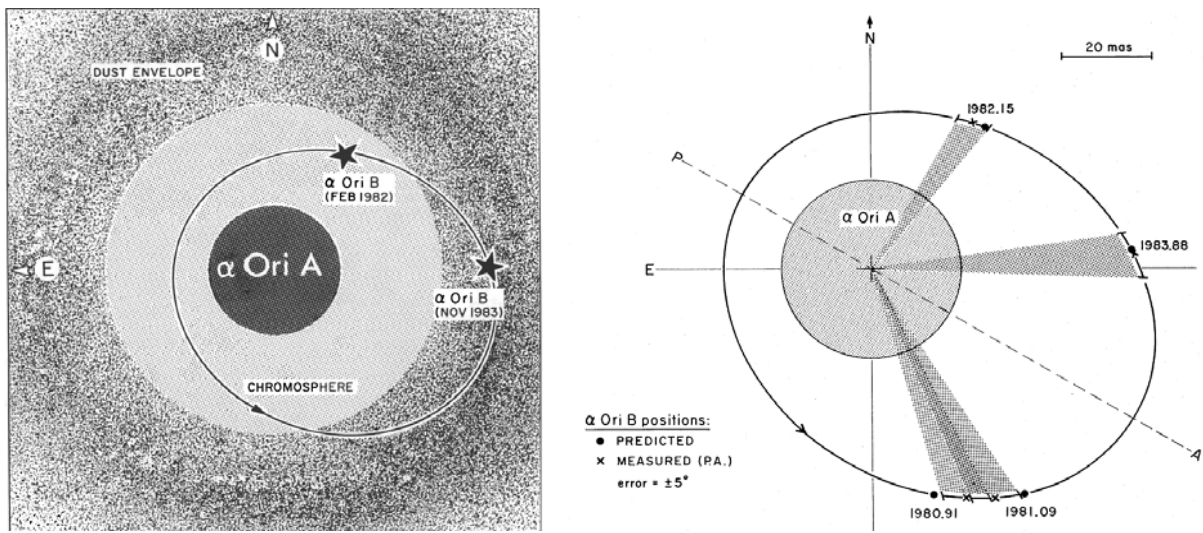

Fig. 7 The multiple system of Betelgeuse, with its extended chromospheres and the dust envelope around the star, from [8].

**The variability**

Betelgeuse is known as an irregular variable star. Stebbins [7] found in 1931 a 5.781 years period of luminosity fluctuation in the B band. This phenomenon has been reconsidered during the increment of luminosity occurred in December 1983 - February 1984.[8]

Changing of luminosity of 0.1 magnitudes occurred over a month are reported by several authors, and differential photometry has been carried on Betelgeuse compared with Gamma and Phi-2 Ori.[9]

Now the AAVSO website www.aavso.org gathers observations of Betelgeuse made worldwide by amateur astronomers.

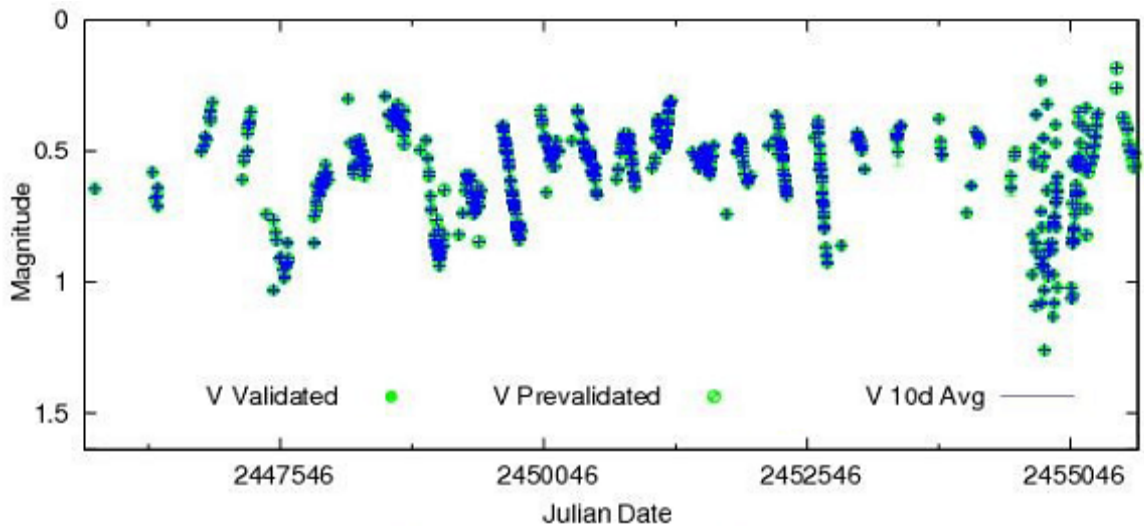

Fig. 8 The luminosity of Betelgeuse in V band, from AAVSO data.